\documentstyle[12pt,psfig]{article}
\newcommand{\be}{\begin{equation}}
\newcommand{\ee}{\end{equation}}
\newcommand{\bea}{\begin{eqnarray}}
\newcommand{\eea}{\end{eqnarray}}

\begin {document} 
\title{Temperature Width and Spin Structure of Superfluid $^{3}He$-$A_{1}$ in 
Aerogel}
\author{G.A. Baramidze and G.A. Kharadze \\ 
\em{Andronikashvili Institute of Physics, Georgian Academy of Sciences}\\
\em{Tamarashvili str. 6, 380077 Tbilisi, Georgia.}}
\maketitle 

%\begin{document}

%\section{\bf Introduction}

\begin{abstract}
The influence of spin-exchange scattering centers on the triplet Cooper pairing is considered to explore 
the behavior of superfluid $^{3}He$ in high porosity aerogel containing  $^{3}He$ atoms localized at 
the surface of silica strands. The homogeneously located and isotropically scattering system of 
spin-polarized ``impurity'' centers is adopted as a  simple model to investigate the contribution of 
spin-exchange scattering chanel for quasiparticles to the formation of non-unitary superfluid 
$A_{1}$-phase in aerogel environment. It is demonstrated that an interference between the potential and 
exchange parts of quasiparticle scattering against spin-polarized ``impurity'' centers can change 
considerably the temperature width and the spin structure of $A_{1}$-phase in aerogel.
\end{abstract}
Among recent achievements in physics of superfluid $ ^{3}He $ the studies of 
the properties of this ordered Fermi liquid in presence of quasiparticle 
scattering medium is of a great importance. This situation is realized for 
liquid  $ ^{3}He $ confined to a high porosity aerogel.

The quasiparticle scattering against silica strands forming skeleton of aerogel
has a profound influence on the properties of such superfluid as liquid 
$ ^{3}He $ in the millikelvin temperature region. The gross effect of a finite mean 
free path $ l $ of quasiparticles is manifested in a sizable suppression of the transition 
temperature of an ordered (superfluid) state, as is expected for a phase with an 
unconventional structure of the order parameter in the momentum space. This behavior 
of superfluid $ ^{3}He $ in aerogel has been observed in a number of experiments using
 various techniques [1-5]

A more delicate question is about possible rearrangement of the phase diagram of 
superfluid $ ^{3}He $ in presence of quasiparticle scattering medium [1-10]. In bulk
superfluid $ ^{3}He $ the isotropic $ B $-phase is a favorable one in the sense that 
in major part of the $ (P,T) $ phase diagram (in zero magnetic field) it appears as 
an equilibrium superfluid state. Only at sufficiently high pressures (above the 
polycritical value $ P_{c0} \simeq 21 $ bars) and at not too low temperatures an 
anisotropic $ A $-phase is a preferable equilibrium state due to so called 
strong-coupling effects [11] which take into account the inverse action of the ordering
on the Cooper pairing interaction between quasiparticles.

In terms of $\beta_{i} $ $(i=1,2, \ldots 5) $ coefficients which appear in the expansion
of the free energy of superfluid  $ ^{3}He $ near the transition to a normal state (at 
 $ T_{c0}(p) $) in power series of the order parameter components, the condition of thermodynamical 
stability of $ ^{3}He-A $ is (in what follows $ \beta_{i j \ldots}=\beta_{i}+\beta_{j}+\ldots $)

%%%%%%%%%%%%%%%%%%%%1
\be\label{condition}
\frac{2\beta_{345}}{3\beta_{13}}<1 .
\ee
 %%%%%%%%%%%%%%%%%%%%
This inequality is not satisfied in a weak-coupling approximation where
 
%%%%%%%%%%%%%%%%%%%%2
\be\label{ConditionWeakCoup}
-2\beta_{1}^{wc}=\beta_{2}^{wc}=\beta_{3}^{wc}=\beta_{4}^{wc}=-\beta_{5}^{wc}=
\frac{7\zeta(3)}{120}\frac{N_{F}}{\pi T_{c0}} ,
\ee
%%%%%%%%%%%%%%%%%%%%
and $ N_{F}$ stands for the quasiparticle density of states (DOS) at the Fermi level.

On introducing pressure-dependent strong-coupling corrections $\delta\beta_{i}^{sc} $
( $\beta_{i}=\beta_{i}^{wc}+\delta\beta_{i}^{sc}$) it can be shown that criterion of
the priority of the $ A $-phase over the $ B $-phase reduces to 

%%%%%%%%%%%%%%%%%%%%3
\be\label{CriterionA>B}
\delta^{\prime}_{sc}(P)>1/4 ,
\ee 
%%%%%%%%%%%%%%%%%%%%
where a dimensionless strong-coupling parameter $ \delta^{\prime}_{sc}(P) $ is 
defined according to an equation

%%%%%%%%%%%%%%%%%%%%4
\be\label{StrCoupParameter}
 \delta^{\prime}_{sc}=1-\frac{\beta_{345}}{2\beta_{13}}.
\ee
%%%%%%%%%%%%%%%%%%%%

It should be noted that along with $ \delta^{\prime}_{sc} $, which
is useful to describe an interplay between $A$- and $B$-phases, some other strong-coupling
parameters can be introduced in an appropriate way. On imposing an external magnetic field 
$^{3}He$-$A_{1}$, characterized by Cooper pairing in a single equal-spin-projection state, is 
stabilized in the vicinity of $ T_{c0} $.  $ A_{1} $-phase appears as an equilibrium state 
below $T_{c1}>T_{c0} $ and extends down to $ T_{c2}<T_{c0} $. As is well known, the 
$A_{1}-A_{2}$ splitting asymmetry ratio

%%%%%%%%%%%%%%%%%%%%5
\be\label{SplittingAsym}
r=\frac{T_{c1}-T_{c0}}{T_{c0}-T_{c2}}=-\frac{\beta_{5}}{\beta_{245}} .
\ee
%%%%%%%%%%%%%%%%%%%
In the weak-coupling approximation $ r=1$ so that the critical temperatures
$ T_{c1} $ and $T_{c2} $ should be positioned symmetrically with respect to 
$ T_{c0} $. In reality a sizable asymmetry of the $ A_{1}-A_{2} $ splitting 
has been observed experimentally [12,13], which is due to strong-coupling effects.
In this case it is convenient to introduce another strong-coupling 
parameter $ \delta^{\prime\prime}_{sc} $ defined by an equation

%%%%%%%%%%%%%%%%%%%%%%%%%%%%%6
\be\label{StrCoupParameter2}
-\frac{\beta_{5}}{\beta_{245}}=\frac{1+ \delta^{\prime\prime}_{sc}/2}   
{1- \delta^{\prime\prime}_{sc}/2}  ,
\ee
%%%%%%%%%%%%%%%%%%%%%%%%%%%%%
or alternatively as $\delta^{\prime\prime}_{sc}=-2\left(1+2\beta_{5}/\beta_{24}\right)$

The phenomenological parameters $ \delta^{\prime}_{sc} $ and
$ \delta^{\prime\prime}_{sc} $ describe the role of the strong-coupling 
effects relative to a weak-coupling contribution $ |\beta^{wc}_{1}| $. 
Generally spiking  $ \delta^{\prime}_{sc}\not=\delta^{\prime\prime}_{sc} $.
In a simple (static) paramagnon model [14] $ \delta\beta^{sc}_{1}=\delta\beta^{sc}_{3}=0 $,
$ \delta\beta^{sc}_{2}=-\delta\beta^{sc}_{4}=-\delta\beta^{sc}_{5}=\delta\beta_{sc} $ and
within this crude approximation  $ \delta^{\prime}_{sc}= \delta^{\prime\prime}_{sc} $ (here
$ \delta\beta_{sc} $ describes the contribution to the strong-coupling effects stemming from
an attractive interaction between quasiparticles via the exchange of magnetic excitations, the
retardation being discarded).

The full temperature width $\Delta T =T_{c1}-T_{c2} $ of the $ A_{1}$-phase in bulk $^{3}He$
is linear in the magnetic field strength ( at list up to $ 10$ T [12]) and is proportional to the 
Ambegaokar-Mermin coefficient $\eta$ [15]. In bulk $^{3}He$ $\eta \not=0 $ due to a small 
particle-hole asymmetry of DOS near the Fermi level. 
For $\Delta T $ we have 

%%%%%%%%%%%%%%%%%%7
\be\label{DeltaT}
\frac{\Delta T}{T_{c0}}=\frac{2\eta h}{1+\delta^{\prime\prime}_{sc}/2} ,
\ee
%%%%%%%%%%%%%%%%%%
where $ h=\gamma H/2T_{c0} $.

Now, on addressing the question of how the phase diagram of superfluid $^{3}He$ could be
modified in aerogel environment, one has to understand in which way the key parameters
$ \delta^{\prime}_{sc} $,  $\delta^{\prime\prime}_{sc}$ and $\eta$, introduced above, react to 
quasiparticle scattering events.

The strong-coupling parameter $\delta^{\prime}_{sc}$ defines, according to Eq.(\ref{CriterionA>B}),
the region of thermodynamical preference of the $A$-phase which in bulk superfluid  $ ^{3}He $
is attained at $ P>P_{c0} \simeq 21 $ bars. The recent acoustic studies [4,5] of superfluid 
$^{3}He$  confined to $98\%$   silica aerogel esteblished that in zero magnetic field the 
$B$-phase-like superfluid state near $ T_{c}(P) $ is stabilized at $ P> P_{c0} $ up to the 
melting pressure $P_{m}$. This observation indicates that scattering of quasiparticles against 
spatial irregularities of a porous medium promotes the stability of the $B$-phase at the 
pressures where in bulk superfluid $^{3}He$ the $A$-phase is an equilibrium ordered state. In 
terms of the strong-coupling parameter $\delta^{\prime}_{sc}$ this means that the equality 
$\delta^{\prime}_{sc}=1/4$ is not reached at $P<P_{m}$ in $98\%$ porosity aerogel and the 
polycritical pressure $P_{c}$ in such quasiparticle momentum non-conserving environment is pushed
to an unobservable region ($P>P_{m}$).

This conclusion is supported by theoretical investigations based on so-called homogeneous 
scattering model (HSM) treating the weak-coupling effect [16] and on a simple (static) paramagnon 
model estimating the strong-coupling contribution [17]. According to Ref.17 in the quasiparticle 
scattering medium

%%%%%%%%%%%%%%%8
\be\label{scattering}
\delta^{\prime}_{sc}=R(w_{c})\delta^{\prime}_{sc0} ,
\ee
%%%%%%%%%%%%%%%
where the subscript ``0'' refers to the corresponding value in bulk superfluid $^{3}He$ and 
the ``impurity'' renormalization factor

%%%%%%%%%%%%%%%9
\be\label{Renormfact}
R(w_{c})=a(w_{c})\frac{T_{c}}{T_{c0}}
\ee
%%%%%%%%%%%%%%%
with
%%%%%%%%%%%%%%%10
\be\label{RenormFactCoef}
a(w_{c})=\frac{\psi^{(1)}(1/2+w_{c})}{\psi^{(1)}(1/2)}\cdot
\frac{\psi^{(2)}(1/2)}{\psi^{(2)}(1/2+w_{c})} .
\ee
%%%%%%%%%%%%%%%
Here $\psi^{(m)}(z)$ is the poly-gamma function of $m$-th order, $w_{c}=\Gamma/2\pi T_{c}$,
where the ``impurity'' scattering rate $ \Gamma=v_{F}/2l$,  and the critical temperature 
$ T_{c}$ of the ``dirty'' superfluid $^{3}He$ is found according to the Abriksov-Gorkov equation
 
%%%%%%%%%%%%%%%11
\be\label{AbrikosovGorkovEq}
\ln\left(\frac{T_{c}}{T_{c0}}\right)+\psi(1/2+w_{c})-\psi(1/2)=0 .
\ee
%%%%%%%%%%%%%%%

The two co-factors in Eq. (\ref{RenormFactCoef}) have opposite behavior as concerns their dependence 
on the scattering parameter $w_{c}:$ $a(w_{c})$ is an increasing function of $w_{c}$ whereas the 
ratio $T_{c}/T_{c0}$ decreases with increasing $w_{c}$. This competition is in favor of 
$T_{c}/T_{c0}$ so that $R(w_{c})<1$ at $w_{c}\not=0$ (for $w_{c} \ll 1$  $R(w_{c})=1-2.56w_{c}$ ).
As a  result strong-coupling parameter $\delta^{\prime}_{sc}$ is suppressed in quasiparticle scattering 
medium thus opening a way to the appearance of a $B$-like superfluid state in the pressure region 
$P>P_{c0}$.

In what follows we concentrate on the $A_{1}-A_{2}$ splitting of superfluid transition in 
aerogel in presence of an external magnetic field. This effect is characterized by the temperature 
width of the $A_{1}$-phase

%%%%%%%%%%%%%%%12
\be\label{TempWidth}
\Delta T= \frac{\eta}{1+\delta^{\prime\prime}_{sc}/2}\gamma H ,
\ee
%%%%%%%%%%%%%%%
and by the field-independent splitting asymmetry ratio
%%%%%%%%%%%%%%%13
\be\label{FAR}
r= \frac{1+\delta^{\prime\prime}_{sc}/2}{1-\delta^{\prime\prime}_{sc}/2} .
\ee
%%%%%%%%%%%%%%%

According to an estimate of strong-coupling effects, mentioned above, it is expected that 
$\delta^{\prime\prime}_{sc}$ is suppressed in aerogel and $r<r_{0}$. On the other hand, the 
$A_{1}$-phase width $\Delta T $ needs a more careful examination. In bulk $^{3}He$  the 
splitting coefficient $\eta_{0}$ stems from a small particle-hole asymmetry of DOS at 
the Fermi level. In the weak-coupling approximation
%%%%%%%%%%%%%%%14
\be\label{ParticleHolAsym}
\eta_{0}=\frac{N^{\prime}}{N_{F}}T_{c0} \ln\left(\frac{2\gamma_{E}}{\pi}\cdot
\frac{\omega_{c}}{T_{c0}}\right)  ,
\ee
%%%%%%%%%%%%%%%
where $ N^{\prime}_{F}$ is the derivative of DOS $N(\varepsilon)$ with respect to the quasiparticle 
excitation energy, $\gamma_{E}$ stands for the Euler constant and $ \omega_{c}$ is a cut-off parameter.

In aerogel environment $\eta_{0}$ is suppressed because of suppression of the critical temperature, 
although this is not the only source of modification of the splitting parameter $\eta$. Below it will 
be shown that more generally

%%%%%%%%%%%%%%%15
\be\label{EtaGen}
\eta=\eta_{0}\frac{T_{c}}{T_{c0}}+\delta\eta  ,
\ee
%%%%%%%%%%%%%
where an extra contribution $ \delta\eta$ is due to the interference part of the spin-exchange
scattering of the quasiparticles against localized $^{3}He$ ``impurity'' atoms adsorbed at the  surface 
of silica strands of aerogel and spin-polarized under the action of an externally imposed magnetic 
field. The presence of such ``frozen'' layers of $^{3}He$ atoms covering aerogel silica strands was
demonstrated in Ref. 18.
 
The spin-triplet Cooper pair condensate is described by an order parameter 
$\vec \Delta(\hat k)$ transforming as a vector on the rotation in spin space. The lowest ordered 
contribution in $\vec \Delta$ to the free energy is proportional to 
$\langle |\vec \Delta|^{2}\rangle $ with brackets $\langle \ldots \rangle $ showing an average 
across the Fermi surface (over the direction of an unity vector $\hat k$ in the momentum space) .
In presence of a magnetic field $\vec H=H \hat h $ a new term 
$ i \langle\vec \Delta \times {\vec \Delta}^{\ast} \rangle \vec H $ appears which contributes 
to the free energy of superfluid $^{3}He$, as long as the particle-hole asymmetry (proportional to
$\eta_{0}$) is taken into account.

In case of spin-triplet Cooper pairing in presence of spin-polarized scattering centers one more
contribution to the free energy emerges proportional to 
$i\langle\vec \Delta \times {\vec \Delta}^{\ast} \rangle \vec S_{T} $, where  $ \vec S_{T}$ is
the thermal average of the localized    ``impurity''   spins [19].  
In order to establish explicitly the quasiparticle spin-exchange scattering contribution 
$\delta\eta$ (as defined by Eq. (\ref{EtaGen})) we adopt the Abrikosov-Gorkov HSM which mimics the 
effects of incoherent scattering of quasiparticles against a system of localized $^{3}He$ atoms 
adsorbed at the surface of aerogel silica strands. The details about HSM of aerogel could be found
in Ref. 20.  In the AG HSM the ``impurity'' scattering interaction is described by 
$2 \times 2 $ matrix

%%%%%%%%%%%%%%16
\be\label{Matrix}
\check U=u_{0} \check I+ u_{ex} \check {\vec \sigma} \vec S .
\ee
%%%%%%%%%%%%%

The rate of potential (spin-independent) part of scattering is characterized by

%%%%%%%%%%%%%%17
\be\label{Potentialrate}
\Gamma=n_{imp}\frac{\sin^{2}\delta_{0}}{\pi N_{F}}=\frac{v_{F}}{2 l}, \qquad 
\tan \delta_{0}=-\pi N_{F} u_{0},
\ee
%%%%%%%%%%%%%%
where $n_{imp}$ stands for an effective concentration of paramagnetic centers and $\delta_{0}$
is an $s$-wave phase shift. In presence of a magnetic field interference part of the scattering 
becomes  operative as long as the polarization of the impurity spins $ \vec S_{T} \not=0$. As a
result the interference scattering rate

%%%%%%%%%%%%%%18
\be\label{Scatteringrate}
\Gamma_{int}=2\pi N_{F} n_{imp} u_{ex}  u_{0}
\ee
%%%%%%%%%%%%%%
appears in the field-dependent contribution to the free energy:

%%%%%%%%%%%%%%%%19
\be\label{FreeEnergy1}
\delta {\cal F}_{SH}=-N_{F}\left[\left(\frac{N^{\prime}_{F}}{N_{F}}\right)
\left(\frac{\gamma H}{2}\right)a_{1}(T_{c})
i \langle\vec \Delta \times {\vec \Delta}^{\ast} \rangle \vec h-
\Gamma_{int} \cos^{4} \delta_{0} a_{2} (T_{c})
i \langle\vec \Delta \times {\vec \Delta}^{\ast} \rangle \vec S_{T_{c}}\right] ,
\ee
%%%%%%%%%%%%%
where
%%%%%%%%%%%%%20%21
\bea\label{coef19}
&&a_{1}=2\pi T \sum_{\omega>0}^{\omega_{c}} \frac{1}{\omega+\Gamma}=
\ln \left( \frac{2\gamma_{E}}{\pi}\frac{\omega_{c}}{T} \right)+\psi(1/2)-\psi(1/2+\Gamma/2\pi T), \\
&&a_{2}=2\pi T \sum_{\omega>0}^{\infty}\frac{1}{\left(\omega+\Gamma\right)^{2}}=
\frac{1}{2\pi T} \psi^{(1)}(1/2+\Gamma/2\pi T)
\eea

Noticing that $\Gamma_{int} \cos^{2} \delta_{0}=(v_{F}/l)
(u_{ex}/u_{0})$ and  adopting a free impurity spin model with 
$S_{T}=\frac{1}{2}\tanh \left(\gamma H/2T\right)$, 
it is concluded that 
quadratic-in-$\vec \Delta$ contribution to the free energy of superfluid $^{3}He$
in aerogel reads as 

%%%%%%%%%%%%%%%%22
\be\label{FreeEnergy2}
{\cal F}_{S}^{(2)}=N_{F}\left(t \langle|\vec \Delta|^{2}\rangle-
\eta h i\langle\vec \Delta \times {\vec \Delta}^{\ast} \rangle \vec h \right)  ,
\ee
%%%%%%%%%%%%%
where $t=(T-T_{c})/T_{c}$, $h=\gamma H/2T_{c}$ and the $A_{1}-A_{2}$ splitting parameter
$\eta$ is given by Eq. (\ref{EtaGen}) with the spin-exchange scattering contribution

%%%%%%%%%%%%%23
\be\label{DeltaEta}
\delta\eta(h)=-\frac{\pi^{2}}{8}\frac{\xi_{c0}}{l}\frac{T_{c0}}{T_{c}}
\frac{u_{ex}}{u_{0}}\frac{\tanh(h)}{h} .
\ee
%%%%%%%%%%%%%
Here  the coherence length $\xi_{c0}=v_{F}/2\pi T_{c0}$ and it is assumed that 
$\cos^{2} \delta_{0} \rightarrow 1/2 $ and $\Gamma \ll 2\pi T_{c}$.

According to Eq. (\ref{TempWidth}) the temperature width of the $A_{1}$-phase in aerogel (relative to 
the bulk value) reads as

%%%%%%%%%%%%%24
\be\label{tempWidthfrac}
\frac{\Delta T}{\left(\Delta T \right)_{0}}=
\frac{|\eta|}{\eta_{0}}\frac{1+\delta^{\prime\prime}_{sc0}/2}{1+\delta^{\prime\prime}_{sc}/2} ,
\ee
%%%%%%%%%%%%

As is evident from Eq. \ref{EtaGen}, in case of $\delta\eta<0$ (realized at $u_{ex}/u_{0}>0$) the 
$A_{1}-A_{2}$ splitting parameter $\eta$ may attain negative values (see below). The measurement of 
temperature width $\Delta T$ do not contain information about the sign of $\eta$ which can be fixed only 
in the experiments where the spin structure ($\uparrow\uparrow$ or $\downarrow\downarrow$) of the 
$A_{1}$-phase Cooper condensate is established (see Ref.21 and citations therein). That is why  
$|\eta|$ stands in Eq. \ref{tempWidthfrac}.

The spin-exchange scattering part $\delta\eta$ can be contribute appreciably to $\Delta T$ at

%%%%%%%%%%%%25
\be\label{noneq}
\frac{\xi_{c0}}{l} \left|\frac{u_{ex}}{u_{0}}\right| \frac{\tanh(h)}{h} 
{^{\sim}_{>}}
\left(\frac{T_{c}}{T_{c0}}\right)^{2}\eta_{0} .
\ee
%%%%%%%%%%
In the low magnetic field case ($\gamma H \ll T_{c}$) this condition is fulfilled at $l=100$ nm, 
$P=15$ bars and $|u_{ex}/u_{0}|=0.1$. For large magnetic fields ($\gamma H \gg T_{c}$) the 
contribution of the spin-exchange scattering part  $\delta\eta$ to the $A_{1}-A_{2}$ splitting 
parameter $\eta$ diminished because of a gradual saturation of the ``impurity'' spin polarization 
$S_{T}$.

\begin{figure}
\vspace{10mm}
\centerline{\psfig{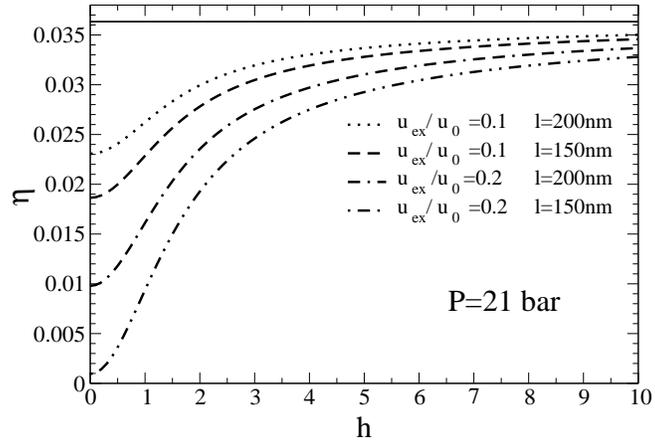}}
\vspace{2mm}
\caption{$\eta$ as a function of $h$, for $P=21$ bar}
\label{fig:etaP=21}
\end{figure}

\begin{figure}
\vspace{10mm}
\centerline{\psfig{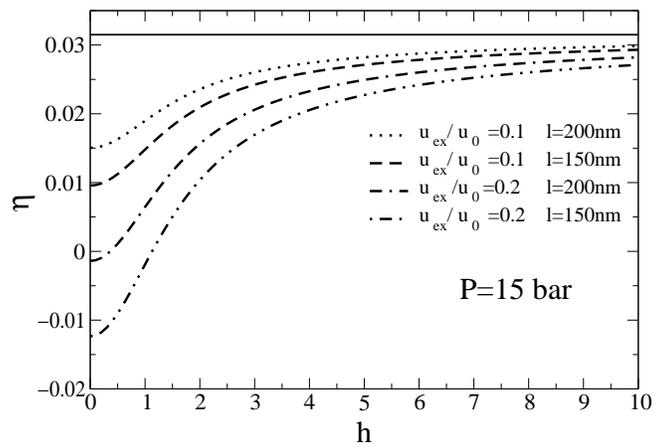}}
\vspace{2mm}
\caption{$\eta$ as a function of $h$, for $P=15$ bar}
\label{fig:etaP=15}
\end{figure}

According to existing experimental data (see Ref. 5) the $A_{1}$-phase temperature width $\Delta T$ 
is suppressed in aerogel environment. Adopting a view that this happens due to the presence of 
quasiparticle spin-exchange scattering contribution $\delta\eta<0$, we concentrate on this possibility. 
Figs.\ref{fig:etaP=21} and \ref{fig:etaP=15} show the dependence $\eta=\eta(h)$ for the pressures 
$P=21$ bars and   $P=15$ bars. It is seen that the spin-polarized scattering centers ($^{3}He$ atoms 
adsorbed at the surface of aerogel silica strands)  suppress considerably the $A_{1}-A_{2}$ splitting 
parameter $\eta$ in relatively low magnetic fields (Fig. \ref{fig:etaP=21}). In the limit of high 
magnetic fields $\eta$ tends to its asymptotic value $(T_{c}/T_{c0})\eta_{0}$. 
Since $|\delta\eta|/\eta_{0}$  is increased on reducing the pressure, the suppression of $\eta$ 
is more pronounced at $P=15$ bars (Fig.\ref{fig:etaP=15}). It is seen that in this situation the 
coefficient $\eta$ can even change sign in low fields.

%%%%%%%%%%%%%%%%%%%%%%%%%%%%%%%%%%%%%%%%%%%%%%%%%%%%%%%%%%%%%%%%%%%%%%%%%%%%%%%%%%%%%%
%%%%%%%%%%%%%%%%%%%%%%%%%%%%%%%%%%%%%%%%%%%%%%%%%%%%%%%%%%%%%%%%%%%%%%%%%%%%%%%%%%%%%%%

%%%%%%%%%%%%%%%%%%%%%%%%%%%%%%%%%%%%%%%%%%%%%%%%%%%%%%%%%%%%%%%%%%%%%%%%%%%%%%%%%%%%%%
%%%%%%%%%%%%%%%%%%%%%%%%%%%%%%%%%%%%%%%%%%%%%%%%%%%%%%%%%%%%%%%%%%%%%%%%%%%%%%%%%%%%%%%%%%

\newpage

\begin{figure}
\vspace{10mm}
\centerline{\psfig{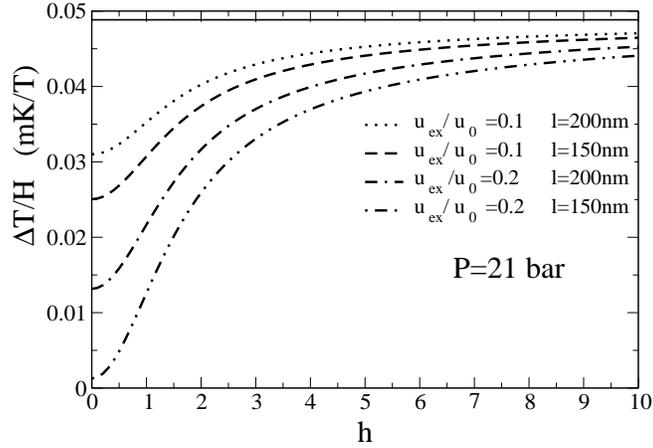}}
\vspace{2mm}
\caption{$\Delta T/H$ as a function of $h$, for $P=21$ bar}
\label{fig:DeltaTP=21}
\end{figure}

\begin{figure}
\vspace{10mm}
\centerline{\psfig{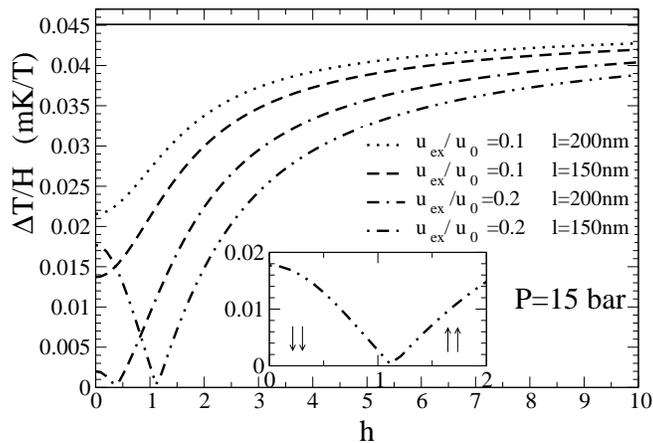}}
\vspace{2mm}
\caption{$\Delta T/H$ as a function of $h$, for $P=15$ bar}
\label{fig:DeltaTP=15}
\end{figure}

In Figs. \ref{fig:DeltaTP=21} and \ref{fig:DeltaTP=15} the values of 

%%%%%%%%%%%%26
\be\label{DeltaTH}
\frac{\Delta T}{H}=\frac{|\eta|}{1+\delta^{\prime\prime}_{sc}/2}\frac{\hbar\gamma}{k_{B}}=
\frac{1.56|\eta|}{1+\delta^{\prime\prime}_{sc}/2}\frac{mK}{T} 
\ee
%%%%%%%%%%
are plotted as a function of $h$ for the pressures $P=21$ bars and $P=15$ bars. A rather peculiar 
situation is expected for $P=15$ bars at $u_{ex}/u_{0}=0.2$ and $l=200$ nm (Fig. \ref{fig:DeltaTP=15}). 
At low magnetic fields (where $\eta<0$) the $A_{1}$-phase with a reversed spin configuration 
$\downarrow\downarrow$ of Cooper pairs is stabilized. On the increase of the magnetic field the 
temperature width of this superfluid state decreases and vanishes at  a field strength for 
which $\eta=0$ (see Fig. \ref{fig:etaP=15}). On further increase of the magnetic field the 
$A_{1}$-phase reappears, this time in  a spin configuration $\uparrow\uparrow$ (appropriate to 
bulk $A_{1}$-phase with $\eta>0$). The reversing of the Cooper pairs spin configuration from 
$\downarrow\downarrow$ to $\uparrow\uparrow$ is shown in the inset of Fig. \ref{fig:DeltaTP=15} at 
$u_{ex}/u_{0}=0.2$ and $l=150$ nm. 

%%%%%%%%%%%%%%%%%%%%%%%%%%%%%%%%%%%%%%%%%%%%%%%%%%%%%%%%%%%%%%%%%%%%%%%%%%%%%%%%%%%%%%
%%%%%%%%%%%%%%%%%%%%%%%%%%%%%%%%%%%%%%%%%%%%%%%%%%%%%%%%%%%%%%%%%%%%%%%%%%%%%%%%%%%%%%%

%%%%%%%%%%%%%%%%%%%%%%%%%%%%%%%%%%%%%%%%%%%%%%%%%%%%%%%%%%%%%%%%%%%%%%%%%%%%%%%%%%%%%%
%%%%%%%%%%%%%%%%%%%%%%%%%%%%%%%%%%%%%%%%%%%%%%%%%%%%%%%%%%%%%%%%%%%%%%%%%%%%%%%%%%%%%%%%%%

In summary, it has been shown that the spin-exchange scattering of quasiparticles against magnetically 
polarized $^{3}He$ atoms adsorbed at the surface of aerogel silica strands can cause substantial 
modification of ($P,T,H$) phase diagram of superfluid $^{3}He$ in the region where non-unitary 
$A_{1}$-phase is stabilized by an externally imposed magnetic field. This effect could be manipulated 
by the variation of the magnetic field strength or by preplating $^{3}He$ in aerogel with some amount 
of  $^{4}He$ atoms which remove paramagnetic scattering centers from silica strands surface. 
\vspace{10mm}

After having completed this article, we learned from Prof. W.P. Halperin about the  Archive preprint 
(cond-mat/0306099) by J.A. Sauls and P. Sharma on the same subject. 
\\
\\
\\
\\
\\
\\
\\
\\
\\
{\large\bf  Acknowledgments}\\
The correspondence with Prof. W.P. Halperin is highly appreciated. We are thankful to Irakli Titvinidze 
for his assistance in numerical constructions. $~~~~~$ This work was partly supported by the 
Grant $N2.17.02$ of Georgian Academy of Sciences.

%%%%%%%%%%%%%%%%%%%%%%%%%%%%%%%%%%%%%%%%%%%%%%%%%%%%%%%%%%%%%%%%%%%%%%%%%%%%%%%%%%%%%%
%\begin{figure}
%\vspace{10mm}
%\centerline{\psfig{file=15neg.eps ,width=85mm,silent=}}
%\vspace{2mm}
%\caption{$\eta$ as a function of $h$, for $P=15$bar}
%\label{fig:etaPN=15}
%\end{figure}
%%%%%%%%%%%%%%%%%%%%%%%%%%%%%%%%%%%%%%%%%%%%%%%%%%%%%%%%%%%%%%%%%%%%%%%%%%%%%%%%%%%%%%%%%%
%%%%%%%%%%%%%%%%%%%%%%%%%%%%%%%%%%%%%%%%%%%%%%%%%%%%%%%%%%%%%%%%%%%%%%%%%%%%%%%%%%%%%%
%\begin{figure}
%\vspace{10mm}
%\centerline{\psfig{file=15DeltaNeg.eps ,width=85mm,silent=}}
%\vspace{2mm}
%\caption{$\Delta T/H$ as a function of $h$, for $P=15$bar}
%\label{fig:DeltaTPN=15}
%\end{figure}
%%%%%%%%%%%%%%%%%%%%%%%%%%%%%%%%%%%%%%%%%%%%%%%%%%%%%%%%%%%%%%%%%%%%%%%%%%%%%%%%%%%%%%%%%%

%%%%%%%%%%%%%%%%%%%%%%%%%%%%%%%%%%%%%%%%%%%%%%%%%%%%%%%%%%%%%%%%%%%%%%%%%%%%%%%%%%%%%%
%\begin{figure}
%\vspace{10mm}
%\centerline{\psfig{file=21neg.eps ,width=85mm,silent=}}
%\vspace{2mm}
%\caption{$\eta$ as a function of $h$, for $P=21$bar}
%\label{figfig:etaPN=21:}
%\end{figure}
%%%%%%%%%%%%%%%%%%%%%%%%%%%%%%%%%%%%%%%%%%%%%%%%%%%%%%%%%%%%%%%%%%%%%%%%%%%%%%%%%%%%%%%
%%%%%%%%%%%%%%%%%%%%%%%%%%%%%%%%%%%%%%%%%%%%%%%%%%%%%%%%%%%%%%%%%%%%%%%%%%%%%%%%%%%%%%
%\begin{figure}
%\vspace{10mm}
%\centerline{\psfig{file=21DeltaNeg.eps ,width=85mm,silent=}}
%\vspace{2mm}
%\caption{$\Delta T/H$ as a function of $h$, for $P=21$bar}
%\label{figfig:DeltaTPN=21:}
%\end{figure}
%%%%%%%%%%%%%%%%%%%%%%%%%%%%%%%%%%%%%%%%%%%%%%%%%%%%%%%%%%%%%%%%%%%%%%%%%%%%%%%%%%%%%%%

%\newpage

%%%%%%%%%%%%%%%%%%%%%%%%%%%%%%%%%%%%%%%%%%%%%%%

%%%%%%%%%%%%%%%%%%%%%%%%%%%%%%%%%%%%%%%%%%%%%%%%%%

\end{document}